\documentclass[12pt]{article} 
\usepackage{amsmath, amssymb, amscd}
\usepackage{graphicx}
\usepackage{epstopdf}

\newcommand{\cell}[1]{\makebox[1.1cm]{#1}}

\begin{document}

\title{Randomized Cellular Automata}
\author{Nino Boccara\\
Department of Physics\\
University of Illinois at Chicago\\
and\\
DRECAM-SPEC, CEN Saclay\\
91191 Gif-sur-Yvette, France\\
\texttt{boccara@uic.edu}}

\date{}

\maketitle

\section*{Abstract} We define and study a few properties of a class of random automata networks. While regular finite one-dimensional cellular automata are defined on periodic lattices,  these automata networks, called randomized cellular automata,  are defined on random directed graphs with constant out-degrees and evolve according to cellular automaton rules. For some families of rules, a few typical 
\emph{a priori} unexpected results are presented.   

\section{Introduction}
The purpose of this paper is to define and study a few properties of a class of random automata networks. Automata networks are discrete dynamical systems which may be defined as follows. Let $G=(V,E)$ be a \emph{directed} graph, where $V$ is the set of vertices and $E$ the set of edges, i.e., a set of ordered pairs of vertices. An automata network on $V$ is a triple $(G,\mathcal{Q},\{f_i\mid i\in V\})$, where $G$ is a directed graph on $V$, $\mathcal{Q}$ a finite set of states usually equal to $\{0,1,2,\ldots, q-1\}$, and $f_i:\mathcal{Q}^{|U_i|}\mapsto\mathcal{Q}$ a mapping called the \emph{evolution rule} associated to vertex $i$. $U_i=\{j\in V \mid(i,j)\in E\}$ is the neighborhood of $i$, i.e., the set of vertices connected to $i$, and $|U_i|$ is the number of vertices belonging to $U_i$. 

If $V=\mathbb{Z}_L$, the set of integers modulo $L$, $U_i = (i-r_\ell, i-r_\ell+1,\ldots,i-1,i,i+1,\ldots,i+r_r)$,  i.e., neighborhoods are translation-invariant, and the mappings $f_i$ are not site-dependent and all equal to $f$, the corresponding automata network is a finite one-dimensional $n$-input \emph{cellular automaton} (CA) of size $L$, where $r_\ell$ and $r_r$ are, respectively, the \emph{left} and \emph{right radii} of rule $f$ and $n=r_\ell+r_r+1$. The global state of such a CA is represented by a \emph{configuration}, i.e., an application $c: \mathbb{Z}_L\mapsto\mathcal{Q}$. The space $\mathcal{C}$ of all configurations is, therefore, equal to $\mathcal{Q}^{ \mathbb{Z}_L}$. If $s(i,t)$ denotes the state of vertex $i$ at time $t$, its state at time $t+1$ is given by
\begin{equation}
s(i,t+1) = f\big(s(i-r_\ell, t), s(i-r_\ell+1,t),\ldots,s(i+r_r,t).
\label{CArule}
\end{equation}
The application $S_t:i\mapsto s(i,t)$ defined on $\mathbb{Z}_L$ is the state of the CA at time $t$. Since the state $S_{t+1}$ is entirely determined by the state $S_t$ at time $t$ and the rule $f$, there exists a mapping $F:\mathcal{C}\mapsto\mathcal{C}$ such that $S_{t+1}=F(S_t)$. That is, if $c:i\mapsto c(i)$ is a configuration, then its image $F(c)$ is the configuration
$$
F(c):i\mapsto f\big(c(i-r_\ell), c(i-r_\ell+1),\ldots,c(i+r_r)\big).
$$
$F$ is referred to as the CA \emph{global rule}. As time $t$ increases, the set $F^t(\mathcal{C})$ evolves towards the \emph{limit set} $\Lambda_F$, i.e., 
$$
\Lambda_F = \lim_{t\to\infty}F^t(\mathcal{C})=\bigcap_{t\geq0}F^t(\mathcal{C}).
$$
$\Lambda_F$ is the maximal $F$-invariant subset of $\mathcal{C}$, i.e., if $X\subseteq\mathcal{C}$ is such that $F(X)=X$, then $X\subset\Lambda_F$. 

On the set of all one-dimensional $q$-state $n$-input CA rules defined on $\mathbb{Z}_L$ the operators $R$ and $C$, respectively called \emph{reflection} and \emph{conjugation} and defined  by
\begin{align*}
R\,f(x_1,x_2,\ldots,x_n) & = f(x_n,x_{n-1},\ldots,x_1)\\
C\,f(x_1,x_2,\ldots,x_n) & = f(q-1-x_1, q-1-x_2,\ldots,q-1-x_n),
\end{align*}
are the generators of a group of transformations that allows to partition the set of rules into equivalence classes. Rules belonging to the same class have similar properties.

The temporal evolution of a one-dimensional CA of size $L$ is represented by its \emph{spatiotemporal pattern}, i.e., the two-dimensional subspace
$$
\{(i,t)\mid i\in\mathbb{Z}_L, t\in\mathbb{N}\},
$$ 
where $\mathbb{N}$ is the set of nonnegative integers.

The simplest CAs are the so-called \emph{elementary cellular automata} in which the finite set of states is $\mathcal{Q}=\{0,1\}$ and the rule's radii are both equal to $1$.  There exist $2^{2^3}=256$ different elementary CA rules $f:\{0,1\}^3\to\{0,1\}$.  The rule of an elementary CA can be specified by its look-up table that gives the image of each of the eight three-vertex neighborhoods. That is, any sequence of eight binary digits specifies an elementary CA rule. Here is an example:

\begin{center}
\begin{tabular}{|c|c|c|c|c|c|c|c|}
\hline
\cell{\tt 111} & \cell{\tt 110} & \cell{\tt 101} & \cell{\tt 100}
& \cell{\tt 011} & \cell{\tt 010} & \cell{\tt 001} & \cell{\tt 000} \\
\hline
\cell{\tt 1} & \cell{\tt 0} & \cell{\tt 1} & \cell{\tt 1} &
\cell{\tt 1} & \cell{\tt 0} & \cell{\tt 0} & \cell{\tt 0}\\
\hline
\end{tabular}
\end{center}

Following Wolfram~\cite{W1983}, a \emph{rule code number} may
be associated with each CA rule. If $\mathcal{Q}=\{0,1\}$,
this code number is the decimal value of the binary sequence of
images. For instance, the code
number of the rule above is 184 since
$$
10111000_2 = 2^7+2^5+2^4+2^3 = 184_{10}.
$$
More generally, the code number $N(f)$
of a one-dimensional $|\mathcal{Q}|$-state $n$-input CA rule
$f$ is defined by
$$
N(f) = \sum_{(x_1,x_2,\ldots,x_n)\in \mathcal{Q}^n}
f(x_1,x_2,\ldots,x_n)|\mathcal{Q}|^{|\mathcal{Q}|^{n-1}x_1+
|\mathcal{Q}|^{n-2}x_2+\cdots+|\mathcal{Q}|^0x_n}.
$$

\begin{figure}[ht]
\centering\includegraphics[scale=0.8]{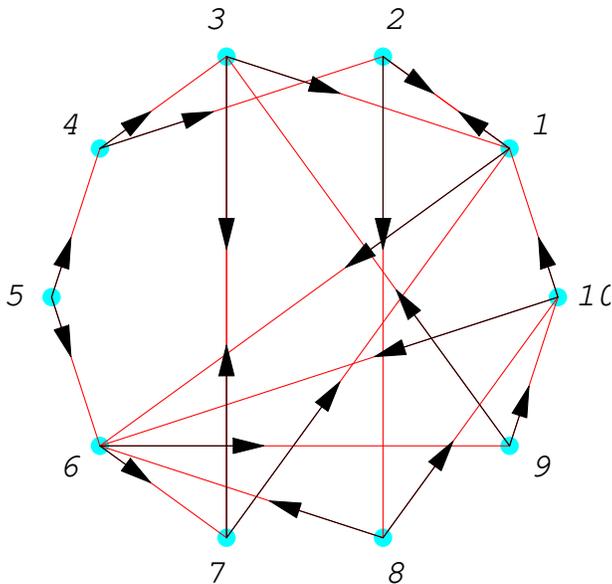}
\caption{\label{fig:randomizedGraph}\textit{A randomized directed graph: $V=\mathbb{Z}_{10}$, $n=3$, $p=0.6$. The links $\{(i,i)\mid i\in V\}$ have not been represented.}}
\end{figure}

If, for all $i\in\mathbb{Z}_L$, each vertex of the neighborhood $U_i$  except $i$ is replaced with a probability $p$ by a randomly selected vertex in $\mathbb{Z}_L$, in such a way that the vertices of every resulting neighborhood are different, we call the corresponding automata network a \emph{randomized cellular automaton} (RCA). Neighborhoods $\{U_i\mid i\in \mathbb{Z}_L\}$ are no more translation-invariant. An example of randomized directed graph is represented in Figure~\ref{fig:randomizedGraph}. 

This process of reconnecting vertices to different ``neighbors''  has been used by Watts and Strogatz  to build up the first example of small-world network~\cite{WS1998}. In the case of a RCA there is, however, an important constraint.  The out-degrees 
$\{d_{\rm out}(i)\mid i\in\mathbb{Z}_L\}$ 
of all the vertices of the randomized directed graph on which is defined the new dynamical system are all equal, that is, for all $i\in\mathbb{Z}_L$,  
$d_{\rm out}(i)=|U_i|-1=r_\ell+r_r$.  Note that the values of $d_{\rm in}$ and $d_{\rm out}$ do not take into account the links $\{(i,i)\mid i\in V\} $.
On the other hand, the in-degree of a given vertex is a random variable $D_{\rm in}$ that can be written as the sum $D_1+D_2$ of two independent binomial random variables. $D_1$ corresponds to the number of original incoming links that have not been reconnected to another vertex during the randomization process, and $D_2$ to the number of new incoming links. 
The generating function 
$G_{\rm in}$ of $D_{\rm in}$ is, therefore, given by
\begin{align*}
G_{\rm in}(z) & = G_1(z) G_2(z)\\ 
\noalign{with}\\
G_1(z) & = (p+(1-p)z)^{r_\ell+r_r}\\
G_2(z) & = \left(\left(1-\frac{1}{L}\right) +\frac{z}{L}\right)^{pL(r_\ell+r_r)}.
\end{align*}
As mentioned above, $L$ is the size of the original one-dimensional CA (i.e., the number of vertices of the graph) supposed to be large. Hence, the average value and the variance of $D_{\rm in}$ are 
\begin{align*}
\langle D_{\rm in}\rangle & = r_\ell+r_r \\
\langle D_{\rm in}^2\rangle - \langle D_{\rm in}\rangle^2 & = 
p(1-p)(r_\ell+r_r)+p\left(1-\frac{1}{L}\right)(r_\ell+r_r). 
\end{align*}
If $L$ is large and $p$ is not too small, i.e., if $pL$ is large, then $D_2$ is well-approximated by a Poisson random variable of parameter $\lambda=r_\ell+r_r$.

If, after randomization, the neighborhood $U_i=(j_{r_\ell},\ldots,j_{-1},i,j_1\ldots,j_{r_r})$,  the state $s(i,t+1)$ of vertex $i$ at time $t+1$ is given by
\begin{equation}
s(i,t+1) = f\big(s(j_{r_\ell},t),\ldots,s(j_{-1},t),s(i,t),s(j_1,t),\ldots,s(j_{r_r},t)\big).
\label{RCArule}
\end{equation}
Translational invariance, which a characteristic property of  regular CAs,  is lost.

\section{Mean-field behavior}
For a given value of the randomizing probability $p$, statistical averages over many different randomized graphs evolving according to a given rule $f$ should tend, as $p$ increases, towards values predicted by the mean-field approximation applied to the corresponding regular CA. We shall see in the next section that this not always the case. In this section we first present a few CA families that exhibit this mean-field behavior. 

\subsection{Surjective CA rules}
A global CA rule $F$ is surjective if any configuration $c\in\mathcal{C}$ has a preimage. This implies $F(\mathcal{C})=\mathcal{C}$.  For a two-state CA, it can be shown~\cite{H1969} that all blocks of zeros and ones of a given length have exactly the same number of preimages. As a consequence, the average value of the density of ones is equal to $\frac{1}{2}$. Since this result is correctly predicted by the mean-field approximation,  the randomization process should have no effect on the value of the average density of ones which should be equal to $\frac{1}{2}$ for all values of $p$.

There exist 30 elementary surjective CA rules. Six of them (rules 15, 51, 85, 170, 204, and 240) are trivial in the sense that their CA rules  $f:\{x_1,x_2,x_3\}\mapsto f(x_1,x_2,x_3)$ are functions of one variable only.
For the remaining 24 rules (rules 30, 45, 60, 75, 86, 89, 90, 101, 102, 105, 106, 120, 135, 149, 150, 153, 154, 165, 166, 169, 180, 195, 210, and 225) we verify that the randomization process, while it destroys the structure of the spatiotemporal pattern, does not change the value of the density of ones, equal to $\frac{1}{2}$ for all values of $p$. It can also be verified that, for not very large CA size (1000), even the statistics of blocks of various lengths also appears to be independent of $p$. 

\subsection{RCA rule 54}
\begin{figure}[ht]
\centering\includegraphics[scale=1.2]{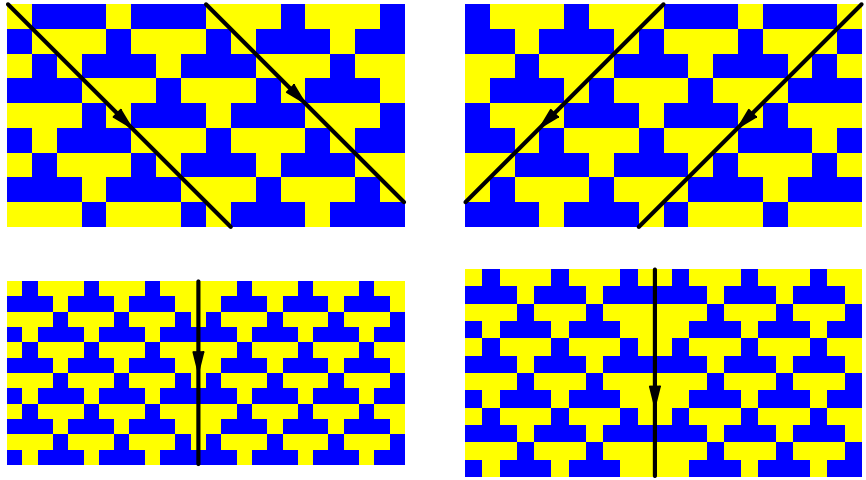}
\caption{\label{fig:ca54defects}\textit{The four different particles found in the spatiotemporal pattern of CA54.}}
\end{figure}

CA rule 54  is defined by the following map
$$
f_{54}(x_1, x_2, x_3) = \begin{cases}
1, & \text{if $(x_1,x_2, x_3)=(0,0,1), (0,1,0),  (1,0,0,)\  \text{or}\  (1,0,1) $,}\\
0, & \text{otherwise}.
\end{cases}
$$
Rule 54 is one of the most complex elementary CA. Its temporal evolution is very slow. It is found that the density of ones tends towards its asymptotic value (exactly $\frac{1}{2}$) as 
$t^{-0.15}$~\cite{BNR1991}. Its spatiotemporal pattern may be viewed as particlelike structures (defects) of 4 different types evolving in a regular background (Figure~\ref{fig:ca54defects} ). These particles present a rich variety of interactions~\cite{BNR1991} which has received  a very nice group-theoretical interpretation due to B. Martin~\cite{M1999}. These particles move in a  periodic background whose space and time periods  are both equal to 4.  This background can be viewed as a square lattice whose elementary cell is the $4\times 4$ tile:
\begin{equation}
\begin{matrix}
0&1&1&1\\
1&0&0&0\\
1&1&0&1\\
0&0&1&0\\
\end{matrix}
\label{b11tile}
\end{equation}

\begin{figure}[h]
\centering\includegraphics[scale=1.2]{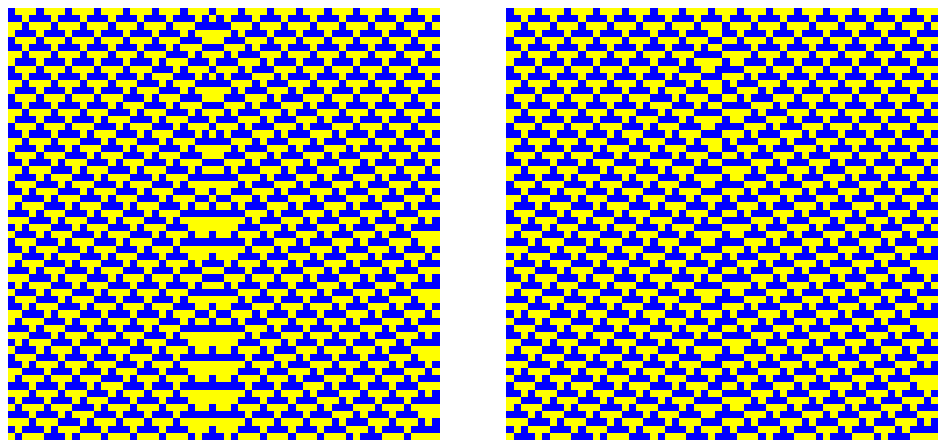}
\caption{\label{fig:radWalls}\textit{Pairs of $g$-particles radiating $w$-particles. }}
\end{figure}

The definition of the $4\times 4$ tile is not unique.  There are actually 8 different $4\times 4$ tiles according to the 8 possible choices of the first row that are~\cite{M1999}:
\begin{equation}
\begin{matrix}
b_{0,0}  &=& (0,0,0,1)  &  & &  b_{0,1} &=& (1,0,1,1) \\
b_{1,0}  &=& (0,0,1,0)  &  & &  b_{1,1} &=& (0,1,1,1) \\
b_{2,0}  &=& (0,1,0,0)  &  & &  b_{2,1} &=& (1,1,1,0)\\
b_{3,0}  &=& (1,0,0,0)  &  & &  b_{3,1} &=& (1,1,0,1)\\
\end{matrix}
\label{martin8tiles}
\end{equation}
Tile~\ref{b11tile} corresponds to background $b_{1,1}$. As shown by Martin, the different particles appear as frontiers between different backgrounds, and can be labeled by the elements of the translation group of the regular background. 

If, starting from a configuration belonging to the limit set,  we rewire one edge (replacing $(i,i+1)$ by $(i,j)$ with $j\neq i+1$), after a few iterations (whose number depends upon the states of the vertices located in the vicinity of vertex $j$) we may obtain very different results. In particular, as shown in Figure~\ref{fig:radWalls}, we may observe rather complicated structures associated to $g$-particles continuously radiating pairs of $w$-particles (on particles' nomenclature refer to~\cite{BNR1991}). Note, on the right figure, the existence of a pair of new defects appearing after one iteration.

Increasing the randomizing probability $p$ eventually destroys the regular spatiotemporal pattern of CA 54. Starting from a random configuration, the structure of the spatiotemporal pattern in the infinite time limit clearly shows that the asymptotic value of the density of ones is equal $\frac{1}{2}$. This result is correctly predicted by the mean-field approximation but, since the evolution of CA 54 is extremely slow, it takes a very large number of iterations to approach this limit. Increasing the probability $p$, notably increases the convergence towards this limit.

\subsection{RCA 18}
For a given value of the randomizing probability $p$, statistical averages over many different randomized graphs evolving according to a given rule $f$ should tend, as $p$ increases, towards values predicted by the mean-field approximation applied to the corresponding regular CA. This is the case of RCA 18.

Rule 18 is defined by the following map
$$
f_{18}(x_1, x_2, x_3) = \begin{cases}
1, & \text{if $(x_1,x_2, x_3)=(0,0,1)\  \text{or}\  (1,0,0)$,}\\
0, & \text{otherwise}.
\end{cases}
$$
The temporal evolution of CA 18 is well understood and rather simple. Configurations belonging to the limit set consist of sequences of zeros of odd lengths separated by isolated ones. The average number of sequences of length $2n+1$ per vertex is equal to $1/2^{n+3}$~\cite{BNR1990}. This implies that, in the limit of large $L$, the fraction of ones of a configuration belonging to the limit set is exactly equal to $\frac{1}{4}$. In the spatiotemporal pattern a sequence of two ones or a sequence of zeros of even length is a \emph{defect}. Since two sequences of zeros of odd lengths separated by two neighboring ones generate, at the next time step, a sequence of zeros of even length, configurations generated by rule 18 contain defects of one type only. During the evolution these defects behave as moving particles. They perform pseudo random walks and annihilate pairwise when they meet. This process has been studied by Grassberger~\cite{G1983} who found that, starting from a random initial configuration, the density of defects decreases as $t^{-1/2}$. If, following~\cite{BNR1991},
we view these defects as particlelike structures evolving in the background represented by the spatiotemporal pattern generated by configurations belonging to the limit set, the pairwise annihilation may be viewed as the annihilation of a particle-antiparticle pair,  the antiparticle, in this case, being identical to the particle.

If, starting from a configuration belonging to the limit set, we rewire only one edge, that is, we replace, for instance, the edge $(i,i+1)$ by the edge $(i,j)$, where $j\notin\{i-1,i,i+1\}$, after a few iterations (whose number depends upon the states of the vertices located in the vicinity of vertex $j$), a pair of defects is created close to site $i$. As a consequence of the structure of the limit set of CA rule 18, it is not possible to create a single defect,  only pairs of defects may be created. Rewiring randomly more vertices creates more defects, some being of a type not found in the spatiotemporal pattern of CA 18, and eventually the structure of this pattern is completely destroyed.  

\begin{figure}[h]
\centering\includegraphics[scale=0.9]{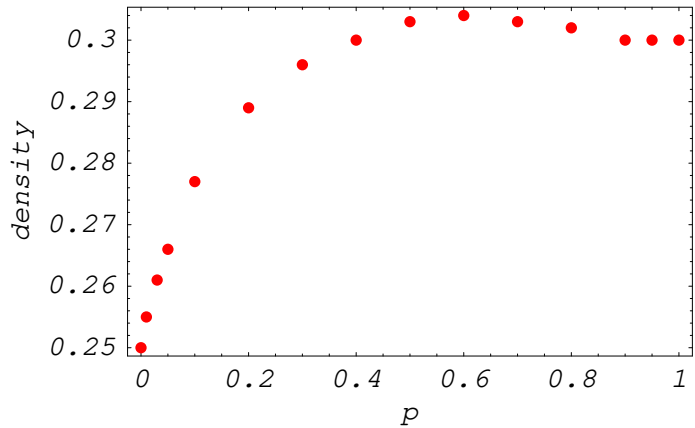}
\caption{\label{fig:rca18Densities}\textit{Densities of ones as a function of the randomizing probability $p$ for RCA 18.}}
\end{figure}

The asymptotic value of the fraction of ones varies as a function of the randomizing probability $p$ (Figure~\ref{fig:rca18Densities} ). For $p=1$, which corresponds to maximum disorder, the density of ones is close to its mean-field value (0.293). 

\section{Number-conserving rules}
A one-dimensional $q$-state $n$-input CA rule $f$ is \textit{number-con\-serving} if, for all cyclic configurations of length $L\ge n$, it satisfies
\begin{multline}
f(x_1,x_2,\ldots,x_{n-1},x_n)+f(x_2,x_3,\ldots,x_n,x_{n+1})+\cdots\\
+f(x_L,x_1\ldots,x_{n-2},x_{n-1})=x_1+x_2+\cdots+x_L.
\label{CRf}
\end{multline}
It can be shown that~\cite{BF2002}

\textbf{Theorem} \textit{A one-dimensional $q$-state $n$-input CA rule $f$ is number-con\-serving if, and only if, for all 
$(x_1,x_2,\ldots,x_n)\in{\mathcal{Q}}^n$, it satisfies}
\begin{align}
f(x_1,x_2,\ldots,x_n) = x_1 + \sum_{k=1}^{n-1}\big(
&f(\underbrace{0,0,\ldots,0}_k,x_2,x_3,\ldots,x_{n-k+1})\notag\\
-&f(\underbrace{0,0,\ldots,0}_k,x_1,x_2,\ldots,x_{n-k})\big),
\label{NScond}
\end{align}

There exist only two number-conserving elementary CA rules, namely rule 184 and its conjugate rule 226. The mean-field map of rule 184 is $\rho\mapsto\rho^3 + 2\rho^2(1-\rho)+\rho(1-\rho)^2=\rho$, where $\rho$ is the density of ones. This result, which is valid for all number-conserving rules, could suggest that the density of ones should not be affected by the randomization process, i.e., the density of ones should, at least in the steady state, remain constant. Our numerical simulations show that this is not the case: the steady state behavior is extremely dependent upon the graph structure.  

\begin{figure}[h]
\centering\includegraphics[scale=0.66]{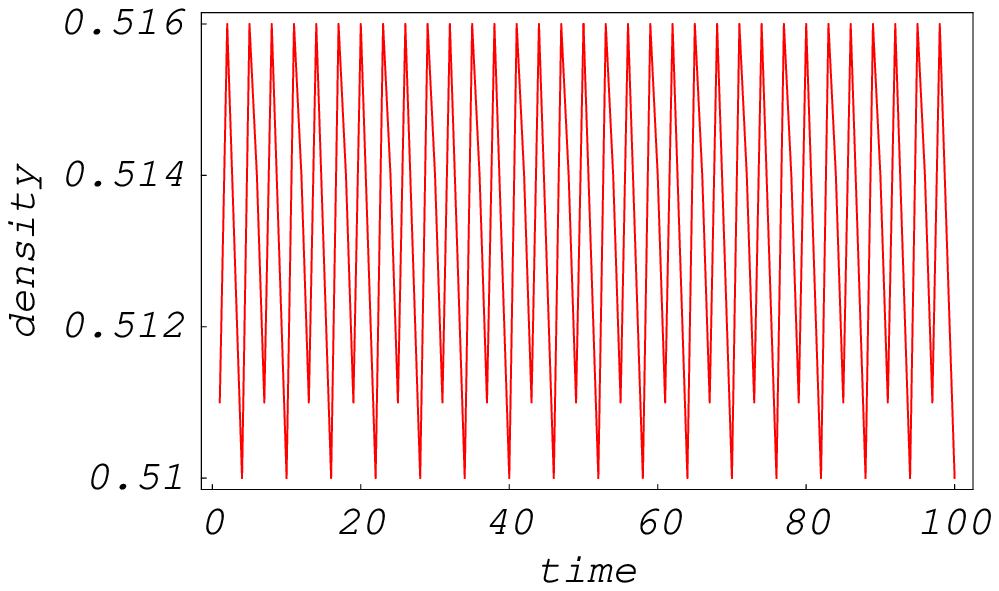}
\vspace{0.5cm}
\centering\includegraphics[scale=0.66]{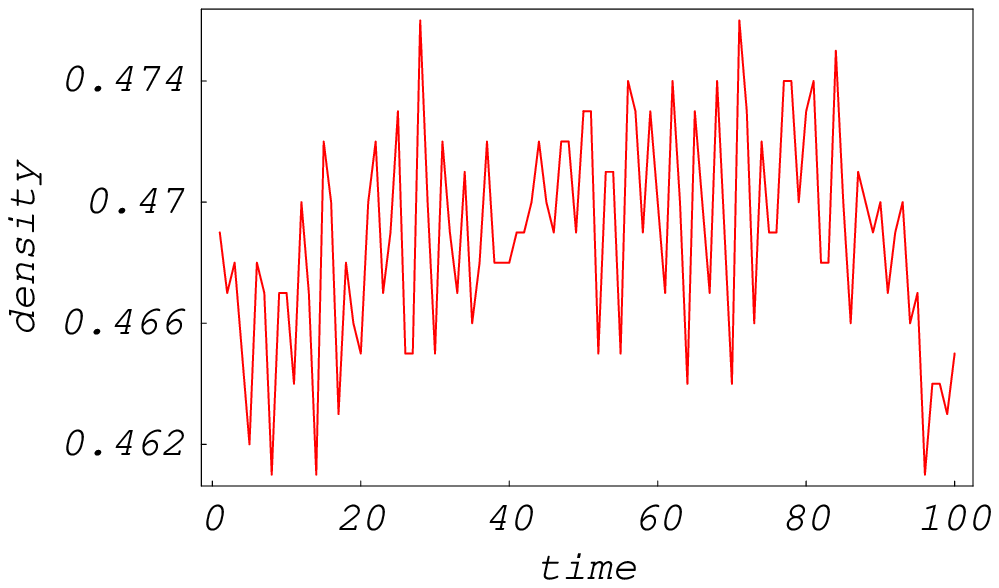}
\caption{\label{fig:rca184p05}\textit{RCA 184: average density of ones as a function of time after a long transient for a randomizing probability $p=0.05$ for two different random structures. Graph size: 1000, initial density: 0.5. Eventually periodic density with period 6 (left figure) and randomly distributed density with mean value equal to 0.469 and a standard deviation equal to  0.00344 (right figure). }}
\end{figure}

For a given value of the randomizing probability $p$, the average density over all vertices becomes either eventually periodic or randomly distributed. If $p$ is small,  the mean value of the the density remains close to the average density of the initial configuration as shown in figure~\ref{fig:rca184p05}.

\begin{figure}[h]
\centering\includegraphics[scale=0.8]{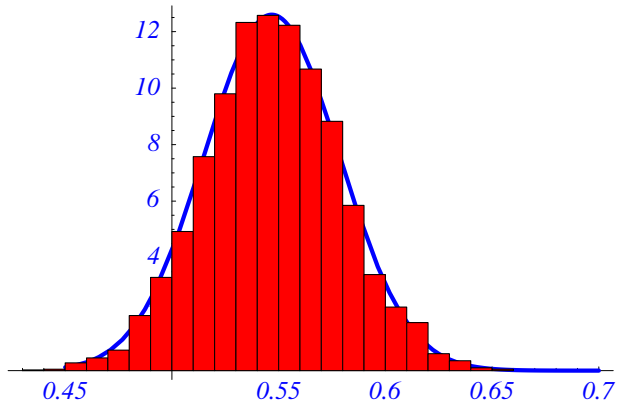}
\caption{\label{fig:rca184p100averageDensityPDF}\textit{RCA 184: Example of a Gaussian probability distribution of the density of ones for $p=1$. Lattice size: 1000, density of the initial configuration: 0.5, mean value: 0.547, standard deviation: 0.0317.}} 
\end{figure}

For larger values of $p$, when the average density is eventually periodic, the period is very sensitive to the graph structure. For example, for $p=1$, changing the structure, periods may takes values such as 2, 6, 17, 61, 304 or 1547, and starting from an initial configuration with an average density of ones equal to 0.7, the mean value of the oscillating density can be as low as 0.13 or as high as 0.97. When the average density is randomly distributed, the probability distribution is also extremely sensitive to the graph structure. As shown in figure~\ref{fig:rca184p100averageDensityPDF} the probability density is sometimes well approximated by a Gaussian distribution, but this is far from being always the case.

\begin{figure}[h]
\centering\includegraphics[scale=0.66]{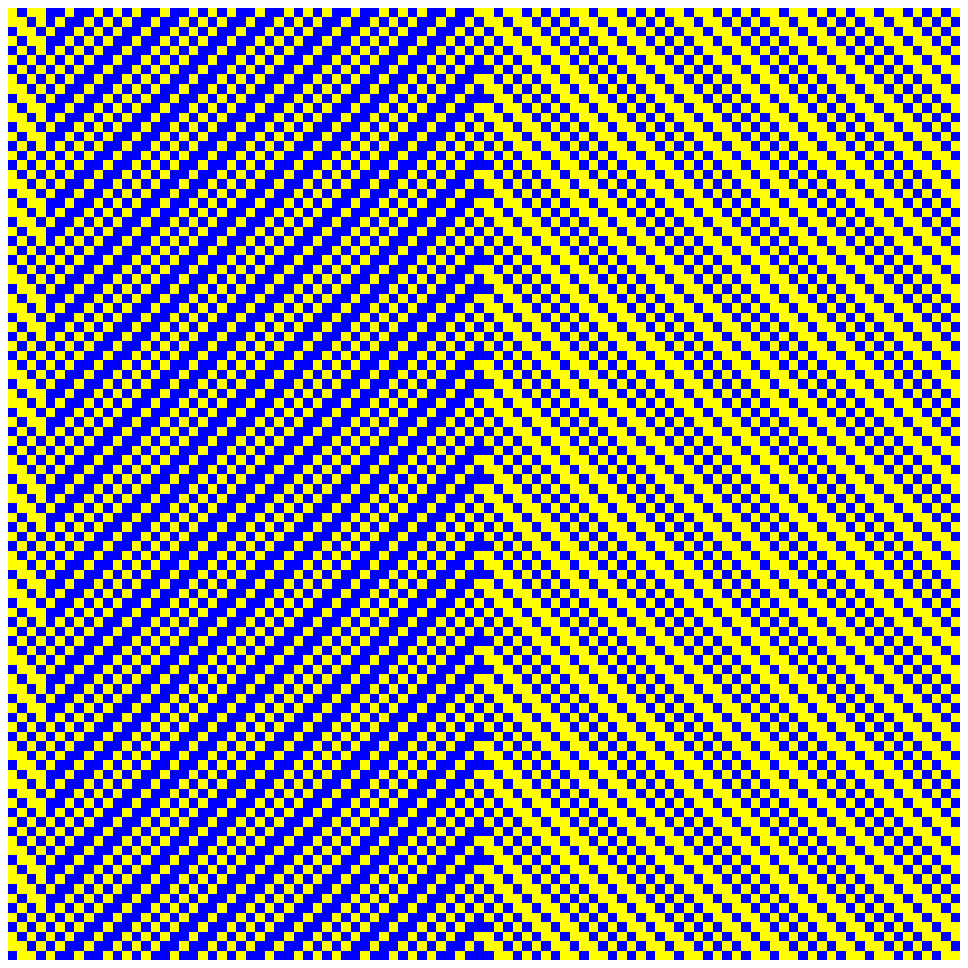}
\vspace{0.5cm}
\centering\includegraphics[scale=0.66]{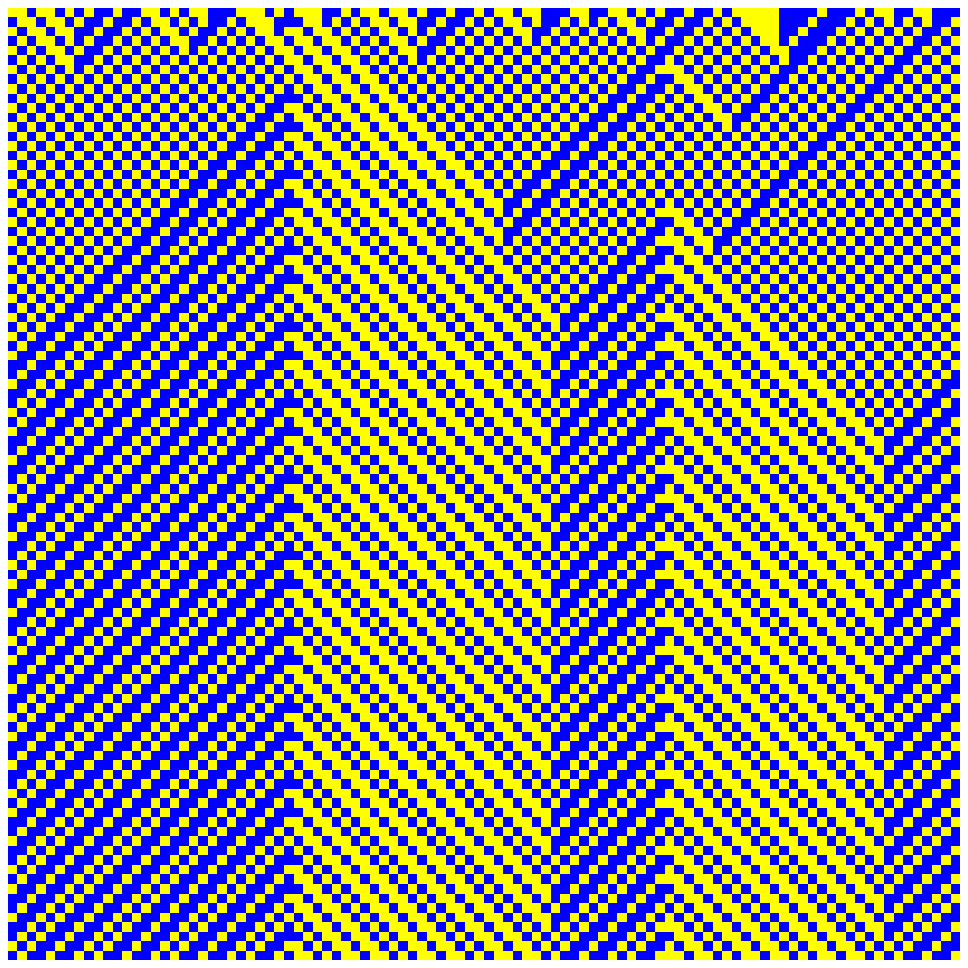}
\caption{\label{fig:fewDefects}\textit{One- and two-defect RCA184, lattice size: 100.}} 
\end{figure}

When the probability $p$ is very small, defects in the spatiotemporal pattern are distant. In the steady state, their numbers remain constant and they interact periodically in time. Hence, the number of active sites is periodic as a result of the spatial periodicity (Figure~\ref{fig:fewDefects}). Increasing  probability $p$ create more defects and their respective distances in a large lattice are not commensurate, the resulting average density of active sites is no more periodic.

Similar results are found for 4-input number-conserving rules except for rules 51148, 56528, 57580, and 62660 which belong to the same equivalence class. These four RCAs exhibit a totally different behavior, they are \emph{eventually number-conserving}, that is, after a finite number of iterations, for all values of the randomizing probability $p$ and all graph structures, the average density becomes time-independent. The final average density depends upon the value of the average density of the initial configuration, the randomizing probability $p$ and, for a given $p$, upon the graph structure.

\section{Conclusion}
We have defined and studied a few properties of a class of random automata networks. The spatial structure of these networks is a directed graph with vertices having a constant out-degrees and random in-degrees resulting from a rewiring process identical to the rewiring process used by  Watts and Strogatz to define the first example of small-world network. When these systems evolve according to cellular automaton rules for different values of the randomizing probability, we find a rich variety of unexpected behaviors, in particular in the case of number-conserving cellular automaton rules.

\end{document}